\documentclass[pre,reprint,superscriptaddress]{revtex4-1}
\usepackage{amsmath}
\usepackage{graphicx}

\begin{document}

\title{Single polymer gating of channels under a solvent gradient}
\author{S Nath}
\affiliation{Department of Physics, Banaras Hindu University, Varanasi 221 005, India}
\author{D P Foster$^*$}
%\email{damien.foster@u-cergy.fr}
\affiliation{Laboratoire de Physique Th\'eorique et Mod\'elisation (CNRS UMR 8089), Universit\'e de Cergy-Pontoise, 2 ave A. Chauvin 95302 Cergy-Pontoise cedex}
\email{Permanent Address: Applied Mathematics Research Centre,}
\email{Coventry University,Coventry, CV1 5FB, UK}
\author{D Giri}
\affiliation{Department of Physics, IIT (BHU), Varanasi 221 005, India}
\author{S Kumar}
\affiliation{Department of Physics, Banaras Hindu University, Varanasi 221 005, India}

\begin{abstract}
We study the effect of a gradient of solvent quality on the coil-globule transition 
for a polymer in a narrow pore.  A simple self-attracting self-avoiding walk model 
of a polymer in solution shows that the variation in the strength of interaction 
across the pore leads  the system to go from one regime (good solvent) to the 
other (poor solvent) across the channel.  This may be thought analogous to 
thermophoresis, where the polymer goes from the hot region
to the cold region under the temperature gradient. 
The behavior of short chains is studied using exact enumeration whilst 
the behavior of long  chains is studied using transfer matrix techniques.
The distribution of the monomer density across the layer suggests that a 
gate-like effect can be created, with potential applications as a sensor.  
\end{abstract}

\maketitle

%\section{Introduction}

In the last few years, much attention has been paid to studying the confinement of 
polymer chains \cite{degennes,binder,grass,sk,dg,pkm}. A confined polymer chain 
 exhibits specific and interesting properties, which have potential 
applications in diverse fields such as steric stabilization of colloidal dispersions, 
behavior of thin films, the adsorption behavior of polymers/ gels, use as surface coatings, understanding the 
behavior of polymers in nanopores etc  \cite{degennes}. However, in most of these studies, 
the surface has been taken as either  neutral or adsorbing and 
statistical mechanics concepts have been applied to the derivation of equilibrium properties. 

When the confining surfaces are put at different temperatures (Fig. 1 a), 
many interesting phenomena take place such as the separation of macromolecules in 
organic solvents \cite{Giddings}, and thermally driven crowding of nucleotides, 
considered as a possible mechanism in 
the molecular evolution of life \cite{Baaske,Budin}. 
Microfluidic applications include colloidal accumulation \cite{Jiang}. 
The temperature gradient acts like an external field, which can drive the system to the 
steady state,  where local thermal equilibrium exists, and local thermodynamic
quantities can be defined. The conductivity, diffusion, and viscosity are just some of the 
physical properties of such a system which have been studied  extensively in the past years. 
These systems are important for the understanding of biological processes such as 
those involved in inflammation, wound healing and cancer metastasis, etc.\cite{keenan}. 
In  these phenomena,  biomolecules move under the chemical gradient. 
The manipulation of a temperature or other gradient is useful  
in many applications which require the control of flow at the nano-metric 
scale, ranging from the control of drug delivery by changing physiological
parameters in the tissue, to the control of flow to a nano-metric 
sensor, where the gradient plays an important role\cite{Adiga}. The presence of a 
polymer in the temperature gradient is then used as a gating device, which depending on the 
gradient will either open or close the pore.

There have been a number of studies, experimental, numerical and theoretical, on 
the behavior of polymer brushes grafted to the walls of  nano-metric channels, 
and their response to changes in solvent quality \cite{Hirokawa,Kokufata,Yoshihiro}, 
electric field \cite{Tanaka}, temperature etc\cite{Jiang,Adiga,Hoffman}. 

\begin{figure}[t]
\includegraphics[width=8cm]{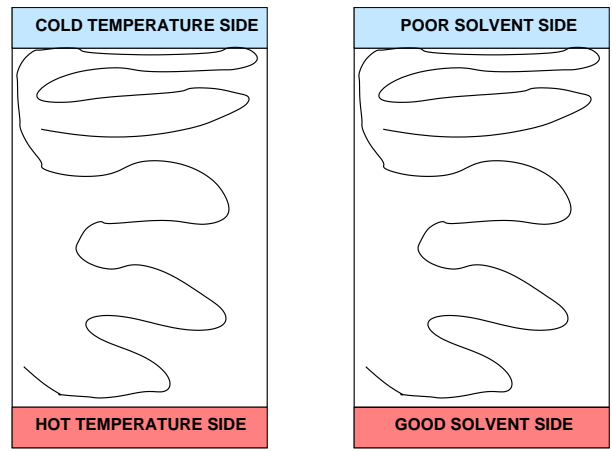}
\caption{(Color online) (a) Schematic representations of the confined surfaces kept at different 
temperatures; (b) corresponding picture in terms of the 
solvent quality. Here, `HOT' corresponds to a `Good solvent', whereas 
`COLD' corresponds to the `Poor solvent'.}\label{model1}
\end{figure}

As the temperature is lowered,  a single polymer chain in solution undergoes a transition from the
coil (high temperature) state to the globule (low temperature) state \cite{degennes}. 
Equilibrium properties of the coil-globule transition are usually derived from the 
partition function, which is the sum over Boltzmann weighted ($e^{\beta \epsilon}$) 
energy conformations \cite{vander,ys}. It is also possible to study the coil-globule transition by 
changing the solvent quality or attractive interaction ($\epsilon > 0$) among monomers.
The solvent is called a good solvent if polymer is found to be in the coil state, whereas 
it is referred as a poor solvent if polymer is in the globule state  \cite{degennes}.
It is not possible to obtain the equilibrium properties of a single 
polymer chain confined between two walls at different temperature, but it is possible
to examine the behavior of a polymer chain to a gradient of solvent quality  across a 
narrow two-dimensional channel. This solvent quality gradient may be thought of as 
resulting from a salt gradient or a temperature gradient (although as discussed above
the system is not strictly in equilibrium, but in the steady state). 
Such a gradient could arise as a result of a difference in the affinity the salt has with 
the top and bottom surfaces of the channel for the salt, or in the case of an ionic salt, 
one could imagine an electric field giving rise to a gradient in the salt concentration.

In this report we present  the behavior of short and  long polymers under such a
solvent gradient. The short polymer case is studied using the exact enumeration method, 
whilst the long polymer case uses the transfer matrices method. We start by presenting 
the basic model and methods used. We  then present the results for both short and 
long chains as a function of the solvent gradient. 
Since we are interested in equilibrium (or pseudo-equilibrium) properties of the polymer, we will 
work with an implicit solvent.  This is valid assuming that the solvent quality gradient affects 
the (effective) nearest-neighbor interaction, and not the chemical potential of the monomers. This 
is the choice often chosen for solvent-quality driven translocation (see for example\cite{WY,LAB}) 
and is reasonable if one thinks of the monomer-monomer interactions having an electro-static 
component, where the salt concentration changes the screening length, for example. As long 
as the screening length is not too large, the approximation of keeping only nearest-neighbor 
interactions is reasonable \cite{nnn}. In the case of the temperature gradient, the change 
of temperature will directly affect the Boltzmann weight associated with the nearest-neighbor 
interaction.

The aim of this paper is to obtain the distribution of density across the channel width as a 
function of the solvent gradient $\Delta \epsilon$.  The  paper ends with a brief discussion on its 
application as a gate for use as a sensor.

%\section{Model and method}

The model used here consists 
of a self-avoiding random walk on the square lattice 
with nearest-neighbor (nn) interactions representing the difference in the affinities 
between the monomers and between monomers and the solvent/salt molecules \cite{vander}. 
In the spirit of such coarse-grained models, the step length, which is the same as the 
lattice spacing in this case, corresponds to a statistical monomer, made up of a number 
of real monomers. 

\begin{figure}[h]
\includegraphics[width=8cm]{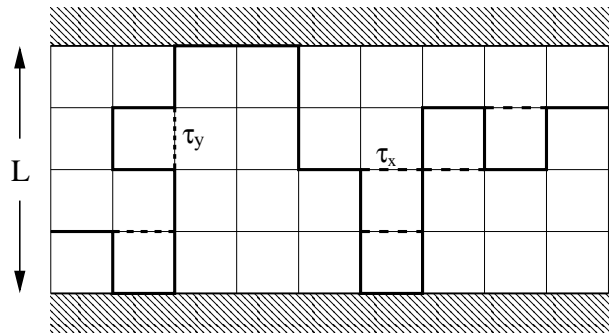}
\caption{Schematic representation of SAW on a square lattice in a confined geometry 
(strip of length $L$). The nearest-neighbor interactions ($\tau_x$ and $\tau_y$) 
are shown by dotted 
lines. For the homogeneous case, the value of $\tau_x$ remains equal to $\tau_y$. However, for 
a solvent gradient, $\tau$ is a function of $y$ and hence $\tau_x \ne \tau_y$. 
See Eqs. 2 \& 3.}
\label{model}
\end{figure}

The thermodynamic properties of a polymer in a solvent of uniform quality are contained in the 
averages of physically 
relevant quantities calculated in the canonical or grand canonical ensembles, hence 
requiring the calculation of the corresponding partition functions. The canonical partition 
function for the homogeneous polymer of length $N$ may be written 
\begin{equation}
Z_N=\sum_{\rm walks} \tau^{N_I}
\end{equation}
where the sum is over self-avoiding walks,
$\tau=\exp(\beta\varepsilon)$ and $N_I$ are the Boltzmann weight for the non-bonded
nearest-neighbor interaction and the number of nearest-neighbor contacts, respectively . 
The nearest-neighbor interaction is attractive, so $\tau>1$. We have chosen a sign convention 
such that the interaction strength $\varepsilon>0$.

The quality of the solvent is changed by varying either the temperature or the 
interaction strength $\varepsilon$. For the two-dimensional square lattice model, the  
$\theta$ temperature, corresponding to the collapse transition of the polymer, for 
the homogeneous model has been found to be at $\beta\varepsilon=0.658\pm0.005$ \cite{fot92}. 
For $\beta\varepsilon$ greater than this value, the polymer is in the bad solvent, whilst 
for a value lower than this, the polymer is in the good solvent.

The solvent quality is now allowed to vary with the distance $y$ from the lower wall, 
and so the interaction strength will be a function of $y$, $\varepsilon(y)$. At the
zeroth layer the nn interaction is zero, whereas at the $L^{th}$ layer, it is 
$\epsilon$. We define  the gradient interaction  $\Delta \epsilon =\frac{\epsilon}{L}$,
%Thus at the first layer, the nn intercation is  $\Delta \epsilon$, whereas at second layer
%it is $2 \Delta \epsilon$ and so on. 
such that the interaction is given by $\varepsilon(y)=y\Delta\varepsilon$ for layer $y$.
To encode this variation in our model, it is 
necessary to consider the weights differently in the $x$ and $y$ directions, 
$\tau_x(y)$ and $\tau_y(y)$. The solvent quality is uniform in the $x$ direction,  and so 

\begin{equation}
\tau_x(y)=\exp\left(\beta\varepsilon(y)\right).
\end{equation}
On the other hand the interaction varies in the $y$ direction, and so we choose 
to define the Boltzmann weight in the $y$ direction with the average interaction strength 
between the two layers involved, such that a nn interaction between layers $y$ 
and $y+1$ will be written as
\begin{equation}
\tau_y(y)=\exp\left(\frac{\beta}{2}\left[\varepsilon(y)+\varepsilon(y+1)\right]\right).
\end{equation}

We study two limiting cases, the case of the short chain and the case of the very long chain. 
In both cases we take the width of the channel to be $L=9$ lattice spaces. 

The short chain case was studied using exact enumeration techniques, where all configurations of 
a chain of length $N=30$ are enumerated, taking into account the $y$ coordinate of any interaction 
and differentiating between interactions in the $x$ and $y$ direction. In practice this was done by 
fixing one end of the chain in one of the layers, enumerating all the configurations and then averaging 
the results over all possible initial layer choices. In this case the canonical partition function (Eq. 1)
for the strip can be written as

\begin{equation}
Z_N^L = \sum_{L,N_I} C(N_I, L) \exp(\beta\varepsilon(y)  N_I),
\end{equation}

The average number of monomers in each layer ($\langle n(y) \rangle$) can be calculated as 
\begin{equation}
\langle n(y) \rangle = \frac{\sum_{L,N_I} n(y) C(N_I, L) \exp(\beta\varepsilon(y) N_I)}{Z_N^L}
\end{equation}

Since in the case of exact enumeration, the number of conformations  increases $\mu^N N^{\gamma-1}$, where
$\mu$ and $\gamma$ are the effective coordination number of the lattice and exponent respectively \cite{vander}
and hence it is not possible to go for a longer chain. However, it is possible to use transfer matrix 
techniques to get long chain behavior.  In this case, it is more convenient to work in the 
grand-canonical ensemble, defining the grand-canonical partition function by
\begin{equation}
{\cal Z}=\sum_N K^N Z_N,
\end{equation}
where $K$ is the fugacity, which controls the average length of the polymer through:
\begin{equation}
\langle N \rangle = \frac{K}{\cal Z}\frac{\partial {\cal Z}}{\partial K}.
\end{equation}
Note that in this context $K$ does not relate to a chemical potential in the usual way, since the polymer is not allowed to exchange monomers with a reservoir of monomers.

\begin{figure}[h]
\includegraphics[width=6cm,clip]{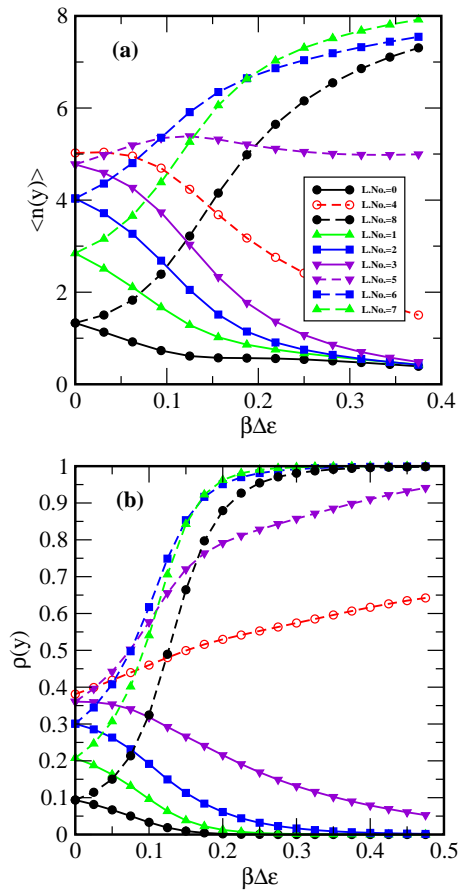}
\caption{(Color online) Distribution of monomers as a function of interaction 
gradient ($\beta\Delta\varepsilon$) for different layers: (a) Using Exact enumeration 
(short chain); (b) Using Transfer matrix (long chain).}
\label{short2}
\end{figure}

The partition function may be calculated exactly on a strip of length $L_x\to\infty$ by defining a 
transfer matrix ${\cal T}$. If periodic boundary conditions are assumed in the $x$-direction, the 
partition function for the strip is given by:
\begin{equation}
{\cal Z}_L=\lim_{L_x\to\infty}{\rm Tr}\left({\cal T}^{L_x}\right).
\end{equation}
The free energy per lattice site, $f$, and the density, $\rho$, may be calculated from the largest eigenvalue, $\lambda_0$
of the transfer matrix:
\begin{eqnarray}
f=\frac{1}{L}\ln\left(\lambda_0\right),\\
\rho= \frac{K}{L\lambda_0}\frac{\partial \lambda_0}{\partial K}.
\end{eqnarray}
where $\lambda_0$ is the largest  eigenvalue of the transfer matrix. Since the solvent quality varies as a 
function of $y$, it is of interest to calculate the local density $\rho(y)$. This may be done by defining 
a local fugacity $K(y)$. The problem is the same as for $\tau$; the density is uniform in $x$ but varies 
in $y$. In calculating $\rho(y)$ we associate half any monomer joining two layers with the layer $y$. 
This is done by defining the fugacity of  a monomer joining layers $y$ and $y+1$ as $\sqrt{K(y)K(y+1)}$. 
The density at layer $y$ is then given by:
\begin{equation}
\rho(y)= \frac{K(y)}{L\lambda_0}\left.\frac{\partial \lambda_0}{\partial K(y)}\right|_{K(y)=K}.
\end{equation}
The fugacity $K$ was chosen such that  the average length of the polymer was taken to infinity setting $\lambda_0(K)=1$.

In Fig.3 (a), we have plotted the distribution of average number of monomers in each layer 
obtained from exact enumeration as function of $\beta \Delta \epsilon$, whereas in Fig. 3(b) we showed the 
density distribution in each layer as a function of $\beta \Delta \epsilon$ for transfer matrix case.  
It is evident from these plots that at zero interaction gradient,
the monomer density is maximum in the middle layer and decreases, when one moves 
towards the surface. 
In order to have the compact structure (collapsed state) the average number of contacts 
needs to be maximized.  This can be achieved only if monomers are distributed (in a compact way) 
near the poor solvent side ($L \approx 8$). In present case 
(chain size N = 30 (31 monomers)), the four layers near $L = 8$ (poor solvent side) are 
populated and, thus the average number of monomers in each layer is
nearly 8, which give rise collapsed state.
It may be noted that the surface here is the hard wall. When
$\Delta \epsilon$ increases, the monomer density shows a clear shift towards  
layer eight, where nearest neighbor interaction is maximum (poor solvent). This is analogous to 
thermophoresis, where the monomer moves away from the hot surface to the 
cold surface \cite{Jiang}. The finite size effect one may notice from Fig. 3a and 3b, where average
number of monomers in layer 4 decreases to zero, whereas for a longer chain it approaches towards
the maximum so that it can have a compact structure.

\begin{figure}[th]
\includegraphics[width=8cm,clip]{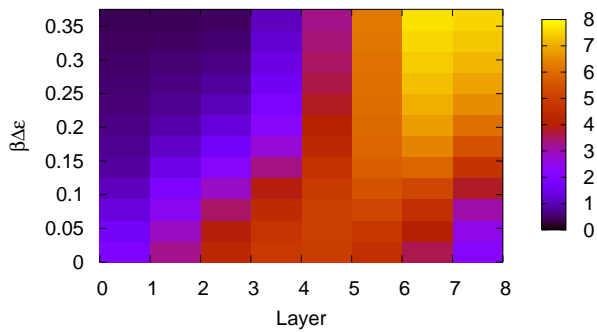}
\caption{(Color online) Distribution of average number of monomers as a function of layer number and 
gradient interaction in case of short chain (using exact enumeration method). The color corresponds 
to the density.
}\label{short}
\end{figure}

\begin{figure}[th]
\includegraphics[width=8cm,clip]{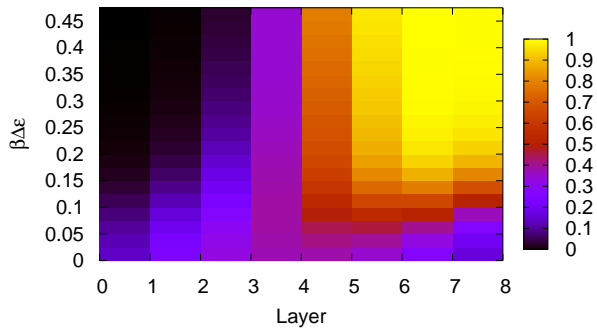}
\caption{(Color online) Same as Fig. 4, but for long chain using the transfer matrix}\label{long}
\end{figure}

\begin{figure}[th]
\includegraphics[width=8cm]{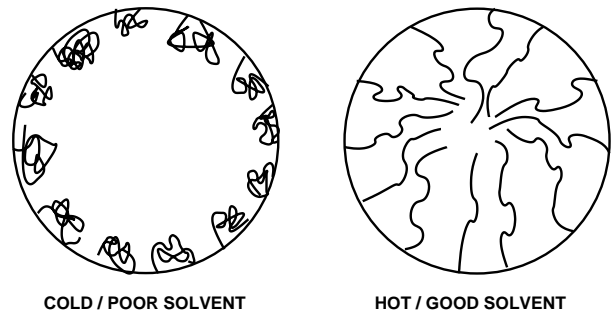}
\caption{Schematic representations of polymers brushes in  a nanocapillary tube: 
(a) Low Temperature (or Poor Solvent condition; (b) High Temperature (or Good Solvent condition)} 
\label{exam1}
\end{figure}

In figures~\ref{short} and~\ref{long} we show the results for the distribution of monomers 
as a function of the layer and interaction gradient. The color corresponds to the density. 
For the short chain,  we plot the average number of monomers in each layer as a function of 
$\beta\Delta\varepsilon$, whilst in the long chain case we plot the density of monomers per 
layer. In each case, the qualitative features are the same: for a good solvent (or 
at high temperatures) the polymer is uniformly distributed across the width of the strip, 
hindering the passage down the strip, whilst at low temperatures, the polymer collapses 
into the region (yellow shaded) where the solvent quality is poor, liberating the good solvent 
portion of the strip i.e. opening up the lower part of the strip. 

\begin{figure}[t]
\includegraphics[width=8cm,clip]{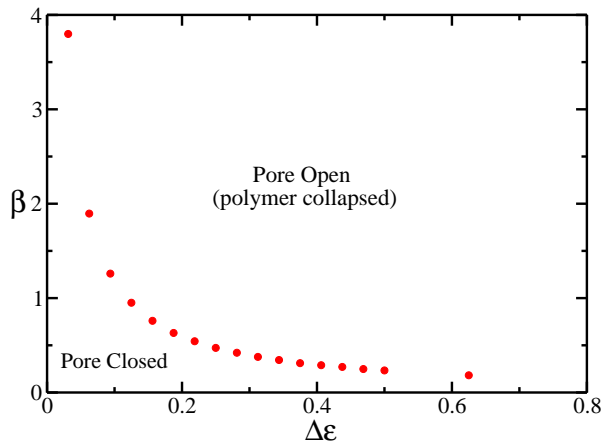}
\caption{(Color online) Phase Diagram ($\beta$ vs $\Delta\varepsilon$) for short polymer chain. 
It shows a transition from closed pore to open pore (polymer collapsed).} \label{kumplot}
\end{figure}

It has been observed that walls coated with polymer brushes act as a valve in a 
nanocaplillary tube \cite{Sevick} under a pressure gradient. Here our studies revealed that
it is possible to design a valve, which can act as a solvent sensor or temperature
sensor. In order to implement it, let us consider a capillary tube grafted with 
many polymers on its surface as shown in Fig. 6. Let us assume the surface of a tube
is at low temperature or tube is filled with a poor solvent. In such a case,
 the polymer will be in the collapsed state and flow of liquid is possible through pore
(Fig.6a). When a good solvent comes inside the capillary tube or the temperature of the
surface changes \cite{Jiang}, there will be a solvent gradient \cite{Yoshihiro} or temperature gradient \cite{Jiang}, which will
swell the polymer and thereby stops the flow of liquid. By suitably adjusting the gradient and 
temperature, it is then possible to gate the pore/capillary by changing one or other 
parameters. Whilst in a narrow pore there is no sharp transition from the swollen 
(good solution) to collapsed (poor solution) behavior, we can derive an estimate of where 
the pore gating will occur by identifying the peak of the density fluctuations as a function 
of the temperature. This is shown as a phase diagram \cite{text}for the short polymers in 
figure~\ref{kumplot}. The ``transition" line lies on the curve $\beta\Delta\varepsilon=0.118(1)$, 
indicating as expected, that the two parameters are interchangeable. For the long polymers 
we find $\beta\Delta\varepsilon=0.140(1)$

Using the lattice model of polymers along with an exact enumeration technique and the transfer matrix 
method, we obtained the exact density distribution of monomers across the strip 
as a function of the gradient of solvent quality.  We have analyzed the chain of different lengths
(finite chain length) and found that the qualitative behaviour remains same and similar
to a longer chain obtained through transfer matrix calculations. This is analogous 
to the distribution of monomers across the strip put at two different temperatures, 
where the usual concepts of equilibrium statistical mechanics fail.
By changing $\beta \Delta \epsilon$, it was shown that gating of the pore may be created  
by either changing solvent quality gradient at fixed temperature or equivalently changing the 
temperature at a fixed (non-zero) gradient. Since one could imagine controlling the solvent quality 
by applying an external field, we proposed a strategy for gating a nanopore. 
The model presented here is simple, but provides the underlying mechanism 
involved in such sensors. 

The financial assistance from the Department of Science \& Technology and Council of Scientific
\& Industrial Research, and Networking Program of University Grants Commissions, New Delhi,  
India is gratefully acknowledged.

%\section{Discussion}


\begin{thebibliography}{99}
\bibitem{nnn} The total interaction may be a combination of screened electrostatic
interactions and van der Waal interactions. Since, electrostatic interactions decay exponentially,
the van der Waals interaction is more important. Equating the electrostatic interaction
with thermal energy, one gets the the characteristic length of the electrostatic interaction
$~ 7-10 \AA$ at room temperature, which is short ranged.
\end{thebibliography}
\end{document}